%% file: MAIN-sample-sigconf.tex
\documentclass[sigconf,natbib=true,review=false]{acmart}

\usepackage{arydshln}
\usepackage{multirow}
\usepackage{times}
\usepackage{latexsym}
\usepackage{graphicx}
\usepackage{booktabs}
\usepackage[T1]{fontenc}
\usepackage[utf8]{inputenc}
\usepackage{microtype}
\usepackage{enumitem}

\AtBeginDocument{%
  \providecommand\BibTeX{{%
    \normalfont B\kern-0.5em{\scshape i\kern-0.25em b}\kern-0.8em\TeX}}}

\setcopyright{rightsretained}
\acmConference[SIGIR eCom'23]{ACM SIGIR Workshop on eCommerce}{July 27, 2023}{ Taipei, Taiwan}
\acmYear{2023}
\copyrightyear{2023}
\makeatletter
\renewcommand\@formatdoi[1]{\ignorespaces}
\makeatother
\acmISBN{}
\begin{document}


\title{Exploring the Viability of Synthetic Query Generation for Relevance Prediction}

\author{Aditi Chaudhary}
\affiliation{%
  \institution{Google Research}
  \city{Mountain View}
  \state{CA}
  \country{USA}
}
\email{aditichaud@google.com}

\author{Karthik Raman}
\affiliation{%
  \institution{Google Research}
  \city{Mountain View}
  \state{CA}
  \country{USA}}
\email{karthikraman@google.com}

\author{Krishna Srinivasan}
\affiliation{%
  \institution{Google Research}
  \city{Mountain View}
  \state{CA}
  \country{USA}}
\email{krishnaps@google.com}

\author{Kazuma Hashimoto}
\affiliation{%
  \institution{Google Research}
  \city{Mountain View}
  \state{CA}
  \country{USA}}
\email{kazumah@google.com}

\author{Mike Bendersky}
\affiliation{%
  \institution{Google Research}
  \city{Mountain View}
  \state{CA}
  \country{USA}}
\email{bemike@google.com}

\author{Marc Najork}
\affiliation{%
  \institution{Google DeepMind}
  \city{Mountain View}
  \state{CA}
  \country{USA}}
\email{najork@google.com }
\renewcommand{\shortauthors}{Chaudhary and Raman, et al.}

\begin{abstract}
Query-document relevance prediction is a critical problem in Information Retrieval systems. 
This problem has increasingly been tackled using (pretrained) transformer-based models which are finetuned using large collections of labeled data. 
However, in specialized domains such as e-commerce and healthcare, the viability of this approach is limited by the dearth of large in-domain data. 
To address this paucity, recent methods leverage these powerful models to generate high-quality task and domain-specific synthetic data. 
Prior work has largely explored \textbf{synthetic data} generation or query generation (QGen) for Question-Answering (QA) and binary (yes/no) relevance prediction, where for instance, the QGen models are given a document, and trained to generate a query relevant to that document. 
However in many problems, we have a more fine-grained notion of relevance than a simple yes/no label. 
Thus, in this work, we conduct a detailed study into how QGen approaches can be leveraged for nuanced relevance prediction. 
We demonstrate that -- contrary to claims from prior works -- current QGen approaches fall short of the more conventional cross-domain transfer-learning approaches. 
Via empirical studies spanning three public e-commerce benchmarks, we identify new shortcomings of existing QGen approaches -- including their inability to distinguish between different grades of relevance. 
To address this, we introduce label-conditioned QGen models which incorporates knowledge about the different relevance. While our experiments demonstrate that these modifications help improve performance of QGen techniques, we also find that QGen approaches struggle to capture the full nuance of the relevance label space and as a result the generated queries are not faithful to the desired relevance label.
\end{abstract}
\maketitle

\input{01-introduction}

\input{02-proposedwork}

\input{03-experimentalsetup}
\input{04-results}
\input{05-relatedwork}
\input{05-conclusion}

\begin{acks}
We would like to thank the anonymous reviewers for the valuable feedback and suggestions.
We would also like to thank Honglei Zhuang and Rolf Jagerman for helping us run and adapt RankT5 model for our experiments.
\end{acks}

%%%%\begin{acks}
%%%%To Robert, for the bagels and explaining CMYK and color spaces.
%%%%\end{acks}

%\clearpage
\bibliographystyle{ACM-Reference-Format}
\bibliography{MAIN-sample-sigconf}
\newpage
\newpage
\input{appendix}


\end{document}

%% file: 01-introduction.tex
\section{Introduction}
\label{sec:intro}
The task of modeling how relevant a document is to a query is among the most central problems in Information Retrieval, and a key component of many IR systems.
The e-commerce domain is no exception, with improved relevance models leading to higher consumer engagement and user satisfaction \cite{Choudhary2022}.
That said, the e-commerce domain offers additional challenges for relevance modeling -- specifically due to its fluidity, with new products appearing every day coupled with the ever-evolving interests of the user base.

The advent of Large Language Models (LLMs) such as GPT~\cite{radford2019language}, T5~\cite{2020t5}, PaLM~\cite{chowdhery2022palm} and LLaMa~\cite{touvron2023llama}, has unlocked new opportunities for potent relevance modeling.
However leveraging LLMs comes with a key requirement: \underline{data}!
As in other IR verticals, e-commerce (relevance) labeled training datasets -- that are large enough to train these LLMs -- are rare\footnote{The one notable exception is the recently released ESCI dataset~\cite{reddy2022shopping} -- which we use and discuss later.}.
The proprietary nature of user logs, coupled with the increasing privacy expectations of users and the exorbitant costs of collecting high-quality relevance ratings, limit the availability of such data.
To tackle this issue, the predominant solution in the IR community has been to leverage large-scale general-purpose IR datasets and perform (zero-shot / few-shot) transfer learning.
In particular the MS-MARCO \cite{DBLP:conf/nips/NguyenRSGTMD16} dataset  -- mined from Bing search logs -- is the largest publicly available dataset (with millions of query-document pairs labeled) and most commonly used to train LLMs to understand query-document relevance.

\begin{figure*}[ht!]
    \centering
    \includegraphics[width=\textwidth]{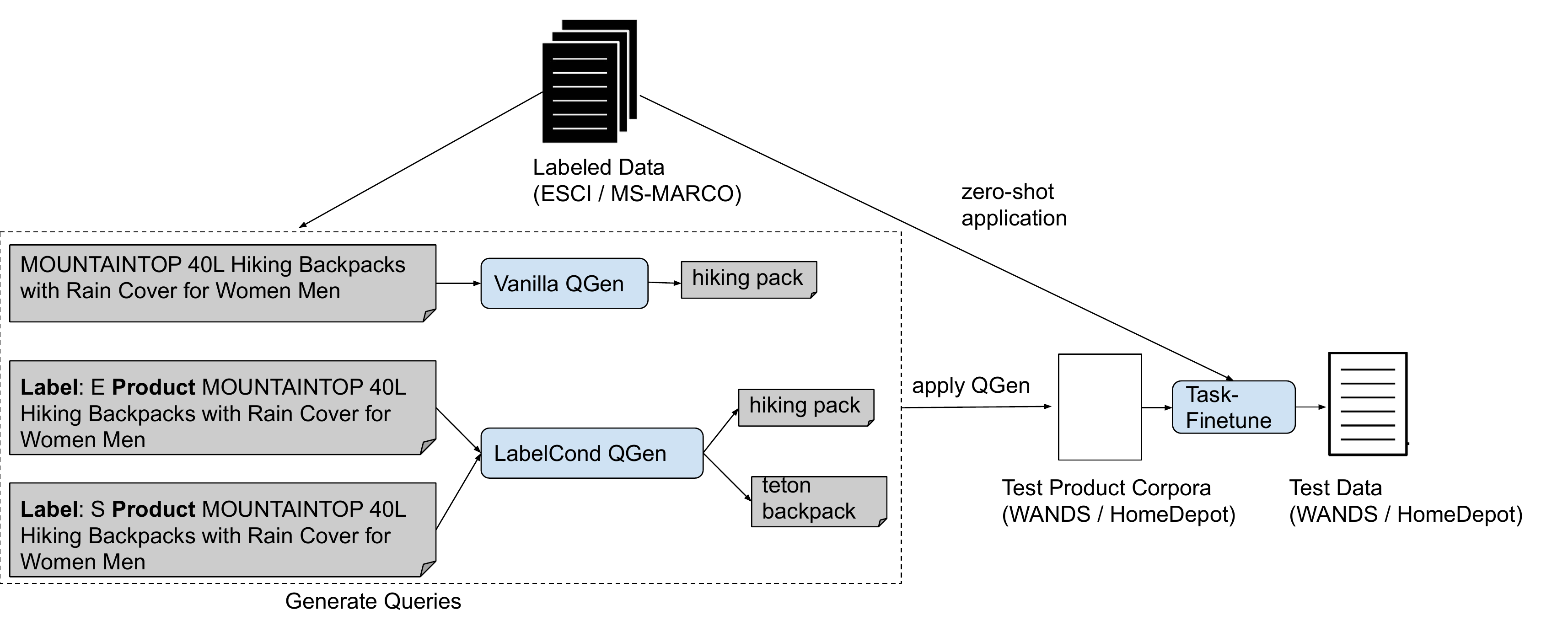}
    \caption{System overview of different approaches we examine for improving relevance prediction on WANDS and HomeDepot.
    \emph{zero-shot application} is a popular transfer learning strategy where labeled data from existing datasets (ESCI (in-domain) or MS-MARCO (out-of-domain) can be used directly to train downstream task model.
    \emph{Generate Queries} approach uses labeled data to train Query Generation (QGen) models -- \emph{vanilla QGen} which most existing works use, where document information, in our case shopping-product information, is used to generate a relevant query and \emph{LabelCond QGen} which we introduce to generate queries by additionally conditioning on all relevance labels. 
    The generated queries are then used to finetune  the downstream model.}
    \label{fig:overview} 
\end{figure*}

Recently, an alternative paradigm has emerged to overcome the lack of query logs -- \textbf{synthetically generated} query logs \emph{i.e.,} Query Generation (QGen). 
Recent works have successfully demonstrated the use of such techniques across different verticals and IR problems, including Question Answering \cite{unlu-menevse-etal-2022-framework}, Passage Ranking \cite{alberti-etal-2019-synthetic} and  Retrieval \cite{ma-etal-2021-zero, dai2022promptagator} -- with some recent results~\cite{dai2022promptagator} even outperforming transfer learning from MS-MARCO.
Beyond improving relevance prediction (the focus of this paper), these synthetically generated query logs can also be used as a substitute for real logs in different IR technologies and problems.
For instance, applications like training query suggestions systems or automatically creating FAQs for consumer-facing applications~\cite{chakrabarty-etal-2022-consistent} could all be performed with such logs.

% Introduce the first contribution: QGen in e-commerce.
Thus our \textbf{first contribution} is providing the first detailed empirical understanding of QGen approaches in the e-commerce domain. 
Using data from three different e-commerce benchmarks, we study performance of the two major families of QGen approaches (finetuning-based vs. prompt-based) popular in the literature.
Our results also demonstrate that models trained using smaller \emph{in-domain} labeled datasets can outperform larger \emph{general-purpose} datasets, thus reinforcing the promise of generating high-quality in-domain synthetic data.

% Second contribution:  Demonstrate that Q-Gen approaches are not always better when model size is comparable. 
Our \textbf{second contribution} involves experiments and analyses that demonstrate that (unlike claims reported in prior works) QGen approaches are outperformed by the more conventional (cross-domain) transfer learning style approaches. 
Via detailed analyses, we identify a set of key reasons (that we have not seen discussed -- or perhaps identified -- in prior works) explaining why QGen approaches fall short. 
For example, we observe that the best existing QGen baseline produces at least one problematic (from the lens of faithfulness / correctness) query for 80+\% of products.

% Third contribution: Nuance QGen works better than vanilla for finetune.
Per our study, a key reason responsible for the shortcomings of existing QGen techniques, is their simplification of the label space. 
More specifically, QGen techniques simplify the problem of query-document (product) relevance into a simple binary one \emph{i.e.,} relevant or not. In fact, most existing approaches only use the relevant query-document pairs, by training the model to produce the associated (relevant) query given the document.
This yes/no binarization is unfortunately a gross over-simplification of the complex relationship between queries and documents. 
For example, TREC relevance judgments are often rated on a 4-point Likert scale. 
Thus ignoring this nuance seems sub-optimal -- as evidenced in our results.
Additionally, as noted in \citet{reddy2022shopping}, nuanced relevance judgements are important for training a high quality product ranker for a better user search experience.
For instance, they define four-class relevance judgements ranging from highly relevant to not relevant. 
A high quality product ranker should be able to rank the highly relevant product over the next relevance class and so on.
Binarizing this would lead to a loss in nuance and thereby the ranking quality.
Thus as our \textbf{third contribution}, we present modifications to both families of existing QGen approaches (finetuning-based and prompt-based) that recognize and leverage the nuance in the relevance label space. Interestingly, while the finetuning variant leads to the overall best QGen models, we find that the prompt-only methods struggle to understand nuance -- indicating potential for future improvements in these pretained prompt models.

%% file: 02-proposedwork.tex
\section{Background: Vanilla QGen}
Powerful transformed-based models like GPT \cite{radford2019language}, T5 \cite{2020t5}, have shown their prowess in generating high-quality text, owing to their ability to attend to even a large context.
These models have now become a starting point for generating synthetic data for training further downstream models.
% More commonly, QGen models have been explored for question generation for improving question-answering (QA) or neural retrieval models.
In this work, we explore two existing paradigms of QGen approaches -- \emph{Finetune-Based} where a QGen model is trained on a subset of training data, and \emph{Prompt-Based} where a large language model (LLM) is leveraged using only few-shot examples.
We refer to these existing approaches as \emph{Vanilla} QGen variants as they use information from only the highest relevance label.
Below, we briefly describe them.
\input{tables/qgen_input_format}

\paragraph{Finetune-Based}
Typically, such a QGen model \cite{lopez2020transformer,alberti-etal-2019-synthetic} is given an input text $d_i$ (e.g. passage or document for question generation)  and is trained to generate a output question $q_i$ which is relevant to that passage or document.
Throughout the paper, the terms `product, document and, passage' are used interchangeably, but they all refer to an input context which is used for generating the query.
Only the relevant query-document pairs from these  datasets (e.g. MS-MARCO, Yahoo Answers, Stack Exchange) are used for training such a QGen model.
The QGen model is then applied to  documents (from the task of interest) to generate synthetic relevant questions.
For training QA models, these new question-document pairs are directly used for data augmentation \cite{alberti-etal-2019-synthetic, unlu-menevse-etal-2022-framework, ko-etal-2020-inquisitive}.
For training neural retrieval models, an additional retriever (e.g. BM25) is then used to retrieve negative documents for every synthetic relevant question \cite{nogueira2019document, ma-etal-2021-zero}.
% In the \emph{answer-aware} paradigm \cite{}, the gold answer $a_i$ is appended to the input $x_i$ to ground the model in generating questions whose answer is $a_i$.

\paragraph{Prompt-Based}
Instead of training a full QGen model, recent works such as \textsc{promptagagor} \cite{dai2022promptagator} and \textsc{InPars} \cite{bonifacio2022inpars} leverage large language models (LLMs) as query generator.
For instance, \textsc{promptagagor} concatenates 8 relevant question-document pairs \\
$\{(q_0, d_0)\cdots(q_7, d_7)\}$ with the target document of interest ($d_t$) and prompts the LLM to generate a new  question ($q_t$) that is relevant to $d_t$.
Then, a retriever is used on the generated new query to  construct hard negatives to train a new model on the downstream ranking task.
\textsc{InPars} uses 3 question-document pairs followed by a BM25 retriever to train a T5-reranker model.
% In addition to this vanilla approach of using only relevant question-document examples, they also propose a variant GBQ (Guided by Bad Questions) which uses provided a bad question with the relevant question

% For classification tasks, the input additionally contains the label information \cite{kumar-etal-2020-data, yang2020generative}, for example, for sentiment analysis task, the input sentence $x_i$ is pre-pended with the label $l_i$ and all examples trained in an auto-regressive fashion.
% During generation, the QGen model is then prompted with a desired label $y_{\text{desired}}$ to generate the input-output pair.
% \cite{kumar-etal-2020-data} test this setup for sentiment analysis (binary relevance) and intent classification (multi-class). 

In this work, we explore the application of these existing QGen approaches to a much harder relevance prediction task, where it has multiple relevance classes which are nuanced as opposed to only binary relevance prediction (e.g in MS-MARCO).
In the next section, we describe our adaptations to the above QGen approaches which conditions on all labels.

\section{Proposed: Label-Conditioned QGen for Fine-Grained Relevance Prediction}
% \aditi{introduce nuanced Qgen by conditioning on desired labels in the input
% Label conditioned QGen adaptaions or nuanced, in Table 3/4 within baselines make in put all QGen techniqeus together}
%and b) there is an inherent ordering within the labels unlike the intent classification case where is no relation between the labels.
As mentioned above, the task of relevance prediction for e-commerce entails -- given a user-issued query and a product, predict the degree of relevance (e.g. highly relevant, partially relevant, irrelevant) between them. 
Consider an example from \autoref{tab:qgeninput}, where we can see the fine-grained difference in queries across different relevance labels for the same ESCI product.
Simply binarizing this task or only considering queries from one relevance label, as done in the above strategies, could risk losing the nuance.
Therefore, we extend the above described vanilla QGen techniques to our nuanced relevance prediction task by conditioning the query generation on the relevance label.
Below we describe our adaptations:
\begin{itemize}
    \item \textbf{Finetune-Based-LabelCond}: we use the entire training portion of the available data and not just the relevant query-document portion, to train the QGen model.
    Specifically, each annotated query-document-label triple is transformed such that the label $l_i$ is prepended to the document $d_i$ and the model is trained to the output the query $q_i$, as shown in \autoref{tab:qgeninput}.
    \item \textbf{Prompt-Based-LabelCond}: we follow \textsc{promptagator} and instead of using all 8 examples of just the relevant label, we use 2 labels per each relevance label, where again, the label is prepended to the respective example as shown in \autoref{tab:qgeninput}.
\end{itemize}
As before, the QGen model is applied to the product corpora of the target domain, which generates query-product examples for all labels, on which a downstream task model is then trained.
In the next section, we describe in detail this entire process.
% Here are all the QGen variants we
% Each annotated query-document-label triple is transformed 
% As we show in \autoref{table}, we find that such a vanilla approach of using the class label helps in improving the downstream performance to some extent but not by a lot because it suffers from a label inconsistency problem i.e. the model often generates the same query for two different levels of relevance suggesting that the model is not respecting the label input.
% The queries produced are relevant to the product but it struggles to discriminate what a query should be for different labels of the same product.
% To address this 

%% file: tables/qgen_input_format.tex
\begin{table*}[t!]
 \resizebox{\textwidth}{!}{
\begin{tabular}{lll}
\toprule
\textbf{Dataset} & \textbf{QGen Input Format} & \textbf{QGen Output Format} \\ 
\midrule
ESCI & \textbf{Label}: E \textbf{Product}: Korean Skin Care K Beauty  .  & \textbf{Query}: vitamin c serum without hyaluronic acid \\
& \textbf{Description}: Seoul Ceuticals CE Ferulic Serum. .. & \\
&  \textbf{Label}: S \textbf{Product}: Korean Skin Care K Beauty  & \textbf{Query}: indie skincare brand \\
& \textbf{Description}: Seoul Ceuticals CE Ferulic Serum. ...  & \\
&  \textbf{Label}: C \textbf{Product}: Korean Skin Care K Beauty   & \textbf{Query}: gundry dark spot diminisher \\
 & \textbf{Description}: Seoul Ceuticals CE Ferulic Serum. .. & \\
&  \textbf{Label}: I \textbf{Product}: Korean Skin Care K Beauty   & \textbf{Query}: victim without a face \\
& \textbf{Description}: Seoul Ceuticals CE Ferulic Serum. .. & \\
\midrule
MS-MARCO & \textbf{Label}: Relevant  \textbf{Document}: Denier (measure): & \textbf{Query}: thread count definition\\
&  Wikis Thread count or threads per inch (TPI) is a measure of  ... & \\
& \textbf{Label}: Irrelevant \textbf{Document}: Denier (measure): & \textbf{Query}: perle cotton thread definition\\
& Wikis Thread count or threads per inch (TPI) is a measure of  ...  & \\

\bottomrule
\end{tabular}
}
\caption{Example QGen input/output formats for the ESCI and MS-MARCO dataset.
The \textbf{bolded words} is the template we use for constructing the input and output text strings -- input is \textbf{Label:} <label> \textbf{Product:} <product title> \textbf{Description:} <product description> and output is \textbf{Query:} <query>.
}
\label{tab:qgeninput}

\vspace{-6mm}
\end{table*}

%% file: 03-experimentalsetup.tex
\section{Experimental Setup}
\label{sec:expt}

We conduct experiments for the zero-shot setting, where we assume that we do not have any training data for our dataset of interest.
We use two e-commerce datasets as our target, namely, WANDS and HomeDepot,  both described below.
These datasets were selected to fulfill the following desiderata -- a) they provide a significantly-sized test sets in the e-commerce domain and has real-world impact, and b) they have fine-grained nuance in the relevance judgements. 
To understand the effectiveness of QGen over the more conventional transfer learning approaches, we compare the cross-domain transfer learning approach (which is non-QGen) with two QGen approaches (vanilla vs label-conditioned).
For the zero-shot cross-domain transfer learning, we train a downstream relevance prediction model on existing datasets, namely, ESCI and MS-MARCO, where 
MS-MARCO is more general-purpose while ESCI is e-commerce focussed, albeit much smaller in size.
The QGen models are similarly trained on ESCI and MS-MARCO, and applied to the two target datasets to create  training data for the downstream task (\autoref{fig:overview}).
We now present the datasets in more detail.

\subsection{Data}
\paragraph{MS-MARCO}
\citet{bajaj2016ms} first introduced the MS MARCO dataset which is constructed from Bing search logs having 8 million passages extracted from general-purpose web documents.
Over the years this dataset has been updated and subsets of it have been used for many shared tasks (e.g. TREC\footnote{\url{https://trec.nist.gov/}}).
In this paper, we use the same MS-MARCO data as used by \citet{zhuang2022rankt5} which comprises of 530,000 queries and a passage corpus of 8 million, each query being annotated with binary relevance judgements (0 for not relevant and 1 for relevant).
Furthermore, \citet{zhuang2022rankt5} retrieve 35 hard negatives for each relevant query and upsample the relevant examples to match the irrelevant examples, we refer the reader to the paper for more details. 
\paragraph{ESCI}
\citet{reddy2022shopping} comprises of 2.6 million manually labeled query-product relevance judgements obtained from the Amazon Search pool.
To the best of our knowledge, this is the largest shopping queries dataset publicly available which comprises of 130k unique queries covering three languages, namely English, Spanish and Japanese. 
The query-product pairs are rated for four relevance labels: \textbf{Exact (E)} when the product is exactly relevant to the query, \textbf{Substitute (S)} when the product is somewhat relevant but it fails to satisfy all requirements of the query (e.g. showing a `red sweater' product for a `blue sweater' query), \textbf{Complement (C)} when the item doesn't satisfy the query but could be used in combination with the query (e.g. showing `hydration pack' for a `hiking bag' query), and \textbf{Irrelevant (I)} when the product is completely irrelevant to the central aspect of the query (e.g. `harry potter book' for a `telescope' query.

\paragraph{WANDS}
\citet{wands} is a product-search relevance dataset released by WayFair \footnote{\url{https://www.wayfair.com/}}, which primarily focusses on home improvement.
It comprises of 233,448 human-annotated relevance judgements comprising of 480 unique queries and 42,994 unique products.
Unlike ESCI, WANDS has been labeled with three relevance labels, namely, \textbf{Exact-Match} where the product fully matches the user query, \textbf{Partial-Match} where the product somewhat matches the query in terms of the target entity but does not satisfy the modifiers, and \textbf{Irrelevant} where the product is not relevant to the user query.
We consider all 233k examples as our test set for evaluation.\footnote{There is no designated train/test split provided so for reproducibility we use the entire data as our test set.}

\paragraph{HomeDepot}
The \textit{Home Depot Product Search Relevance}\footnote{\url{https://www.kaggle.com/c/home-depot-product-search-relevance/}} released by Home Depot\footnote{\url{http://www.homedepot.com.}} retailer, comprises of 73,789 training examples\footnote{The original dataset had 74,068 examples but 279 of those had parsing issues.} and 166k test examples, focussing on home improvement e-commerce.
However, the relevance labels for the test split are not released publicly so we use the entire train portion for our zero-shot evaluation.
These comprise of 54,470 unique products with relevance labels scored between 1 (not relevant) to 3 (highly relevant).

\autoref{tab:qgendatasets} shows some examples for all these datasets and in \autoref{tab:qgenlabels} we describe the statistics for each dataset.

\input{tables/wands_results}
\input{tables/homedepot_results}

\subsection{QGen Setup}\label{sec:setup}
We use the pretrained mT5-XXL \cite{xue-etal-2021-mt5} model (13B parameters) as our starting point for all \textbf{Finetune-Based*} models, which has been trained for 1M steps on a multilingual corpora, which gives our subsequent models the ability to generate in many languages inherently.
For \textbf{Prompt-Based*} models ,we use the same setup as \textsc{promptagator} \cite{dai2022promptagator} which uses FLAN-137B as the large language model (LLM) \cite{wei2021finetuned}.
We use the t5x code base \url{https://github.com/google-research/t5x} to train all models.

\paragraph{Train QGen}
We finetune a QGen model using the training portion of MS-MARCO and ESCI.
We use the same train/dev/test splits as provided with the respective datasets and transform the input/output as shown in \autoref{tab:qgeninput}.
We finetune the QGen model for 100k additional steps with a constant learning rate of 1e-4, Adafactor optimzer, batch size 128, input sequence length 256, target length 32.\footnote{100k steps amount to approximately 8 epochs which we thought were sufficient given the computational and time requirements.}
The best checkpoint for subsequent steps was selected using the performance of BLEU on the validation set.

\paragraph{Apply QGen}
Next, we apply the above trained models to generate query-product pairs on WANDS and HomeDepot.
For the \emph{label-conditioned} models, the input text is a concatenation of the desired label and the product, whereas for its \emph{vanilla} QGen counterparts the input text is simply the product information.
Since \textbf{*-LabelCond} QGen models  have the ability to generate queries for different relevance labels, unlike their vanilla counterparts which only can generate queries for one relevance label, we generate queries for all relevance labels for a given product.
Similar to the training setup, we use input sequence length of 256 with target length to be 32.
% Consider a \textsc{Finetune-Based-AllLabels} QGen model trained on all four labels of ESCI, then for each of the WANDS 42994 unique products, we generate four queries -- one query per each E,S,C,I label, resulting in 171k query-product pairs.
As an additional filtration step, we remove duplicate queries i.e. if the same query is generated for different labels of the same product, we only retain that query-product-label triple which has the highest model probability.\footnote{see \autoref{tab:wands_results_duplicate} for number of duplicate queries.}

\subsection{Evaluate QGen for Utility} \label{sec:utility}
We automatically evaluate the generated synthetic data for its utility to the downstream task.
To do that we evaluate the models trained on above-generated data on the respective test sets.
We  split the resulting filtered QGen data into a train and validation set with a 90:10 ratio such that there is no product overlap across the two sets.
We experiment with two styles of downstream models, \emph{classification} and \emph{ranking}.

\paragraph{classification}
We use a pretrained mt5-XXL based encoder-only model to perform multi-class classification, and  report  NDCG.\footnote{We had also tried an encoder-decoder model for the classification task but found encoder-only model to outperform slightly.}
For ESCI-based QGen models,  this would become a four-class classification task.
We finetune the mt5-encoder for additional 25000 steps with a constant learning rate of 1e-4.\footnote{We tried two other learning rates of 1e-3 and 5e-5 but found them to be under-performing.}
We use a batch size of 64 with input sequence length of 608.
The reason we chose NDCG, a ranking metric, instead of accuracy is because there is a label mismatch between ESCI and WANDS, which helps
avoid an oversimplification by deterministic mapping across the two label sets.
This helps evaluate whether the model is correctly ranking exactly-relevant over partially-relevant over irrelevant.
In order to compute NDCG, we need a relevance score output for each query-document pair.
So, from a downstream model based on the four-way ESCI classification model, we output the prediction probability $p(y_i | x_i)$ where $x_i$ is the concatenation of input query, product title, product description and $y_i$ is the output label (E/S/C/I).
We then compute a final score by taking an expectation of the prediction probabilities by multiplying it with the label weight:
\[ E_i = \sum_{j=\{E,S,C,I\}} p(y_i^j | x_i) * w_j \] 
\[ w = \{ E=3.0, S=2.0, C=1.0, I=0.0\} \]

An astute reader may wonder why we go through the trouble of training a multi-class model as opposed to using a ranking model.
The reason is that we want the ability to generate new queries for different relevance labels. 
This is important for search engines where queries/products that have had more user-clicks are often indexed and served with priority (to avoid latency).
However, rare or new products often are not covered as they do not have any query associated with them, so having the ability to generate queries across relevance labels for such products becomes crucial to increase coverage.
However, for completeness, we do report results from a neural re-ranker and find them to underperforming than the classification model (details in \autoref{sec:results}).

\paragraph{ranking}
In the neural re-ranker setup, we use the RankT5 model \cite{zhuang2022rankt5} which uses T5 encoder with pointwise ranking loss wherein the loss for each query-document pair is independently computed.
The authors train the RankT5 model on MS-MARCO which has binary relevance judgements.
We follow the same modeling setup as them, with the main difference being that we use mt5-XL as our starting point instead of T5-Large, as used by them.\footnote{Due to hardware restrictions we could not train the mt5-XXL model variant with their code setup.}
Input sequence length is 256 with constant learning rate of 1e-4.
This ranking model is used in the \textsc{Finetune-Based} and \textsc{Prompt-Based} QGen baselines to evaluate the downstream performance.
In these baselines, as you recall, the QGen models are trained to generate only relevant queries.
To create training data for the ranking downstream model, we need to create negative query-document pairs as well i.e. documents which are not relevant to a query.
For this, we use a dual-encoder T5-based retriever \cite{ni2021large,t5x-ret-sent}\footnote{We used the mt5-BASE model finetuned with the unsupervised objective proposed by \citet{t5x-ret-pretrain}, based on the t5x-retrieval code base: \url{https://github.com/google-research/t5x_retrieval}.} to retrieve top-35 documents for every generated query.
We use all 35 as our hard negative query-document pairs and upsample the relevant documents to have a equal label distribution and train a RankT5 model.
This model ranks the target query-product pairs so we directly use that to compute NDCG.

Below, we briefly summarize all the model variants we experiment with.

\input{tables/qgen_example}
\subsection{All Model Variants}
First, we describe the baselines which do not use QGen:
\begin{itemize}
    \item \textbf{Random} where for the target datasets the documents for a given query are randomly ranked.
    \item \textbf{zero-shot (ESCI)} where we train a downstream model for multi-class classification on all of the ESCI training data and apply it directly to WANDS and Homedepot test data.
    \item \textbf{zero-shot (MS-MARCO)} where we train a ranking model using RankT5 \cite{zhuang2022rankt5} with the pointwise loss function on the MS-MARCO training data and apply it directly to the WANDS and Homedepot test data.
\end{itemize}
Next, we describe the baselines which use existing QGen approaches:
\begin{itemize}[noitemsep,topsep=2pt]
    \item \textbf{Prompt-Based (ESCI)} where we randomly sample 8 query-product pairs from ESCI having \emph{Exact (E)} relevance label and similar to \textsc{promptagator} prompt FLAN-137B to generate one relevant query for a new WANDS/Homedepot product. 
    For the downstream application, we follow the \textbf{ranking} setup described in \autoref{sec:utility}.
    \item \textbf{FineTune-Based (MSMARCO)} where we finetune the QGen model on only those query-passage pairs from MS-MARCO that have \emph{Relevant} label.
    For every new target product, we generate one relevant query and use the retriever to retrieve 35 documents as negative examples following the \textbf{ranking} setup.
\end{itemize}
Finally, we describe our adaptations of the above QGen approaches:
\begin{itemize}[noitemsep,topsep=2pt]
    \item \textbf{Finetune-Based-LabelCond (ESCI)} where we finetune the QGen model on all ESCI examples, and for every new target product generate queries for all four relevance labels. 
    For the subsequent downstream model, we initialize it with the multi-class classification model trained on all ESCI data (which we had used in our \textbf{zero-shot} setting), and further finetune it on the synthetic data, following the \textbf{classification} setup.
    
    \item \textbf{FineTune-Based-LabelCond (MSMARCO)} where we finetune the QGen model on MS-MARCO examples which and for every new product generate two queries for each of the two relevance labels.
    We use the \textbf{ranking} setup to train the downstream model and initialize it with the MS-MARCO-finetuned-ranking model (used in the \textbf{zero-shot} setting).
    
    \item \textbf{Prompt-Based-LabelCond (ESCI)} where we prompt FLAN-137B with 8 ESCI examples comprising of 2 examples per each relevance label.
    For every new target product, we generate queries for all four relevance labels, and follow \textbf{ranking} setup to train the downstream model.
\end{itemize}

%% file: tables/wands_results.tex
\begin{table*}
\resizebox{\textwidth}{!}{
\begin{tabular}{llllll}
\toprule
\textbf{Type} & \textbf{QGen Model} & \textbf{Downstream Model} &  \textbf{NDCG@5} & \textbf{NDCG@10} & \textbf{NDCG@20}  \\
\midrule
Baseline & - & Random & 0.5776	& 0.5989 & 0.6285\\
 & - & zero-shot (ESCI) & \textbf{0.8902} &	\textbf{0.8927} & \textbf{0.8987}\\
 & - & zero-shot (MS-MARCO) & 0.8642 & 0.8649 & 	0.8681\\
\midrule
Baseline & \textsc{Prompt-Based} (ESCI) & ranking & 0.7062 & 0.7118 &	0.7217\\
(Vanilla QGen) & \textsc{Finetune-Based} (MS-MARCO) & ranking & 0.795	& 0.7984 & 0.8015\\
\midrule
Ours & \textsc{Prompt-Based-LabelCond} (ESCI) & classification & 0.6189	& 0.6355 & 0.6561\\
 & \textsc{Finetune-Based-LabelCond} (ESCI) & classification & 0.847	& 0.8553 & 0.8661\\
 & \textsc{Finetune-Based-LabelCond} (MS-MARCO) & ranking & 0.8213	& 0.8318 & 0.8446\\
\bottomrule
\end{tabular}
}
\caption{Results on WANDS dataset where we report NDCG@5,10,20 (higher the better).
}
\label{tab:wands_results}
\end{table*}

%% file: tables/homedepot_results.tex
\begin{table*}
\resizebox{0.9\textwidth}{!}{
\begin{tabular}{llllll}
\toprule
\textbf{Type} & \textbf{QGen Model} & \textbf{Downstream Model} &  \textbf{NDCG@5} & \textbf{NDCG@10} & \textbf{NDCG@20} \\
\midrule
Baseline & - & Random & 0.8939	& 0.9394 & 0.9433\\
 & - & zero-shot (ESCI) & \textbf{0.9212} & 	\textbf{0.9546} & \textbf{0.9575}\\
 & - & zero-shot (MS-MARCO) & 0.9144	& 0.9509 & 0.9542\\
\midrule
 Baseline & \textsc{Prompt-Based} (ESCI) & ranking & 0.9042	& 0.9445 & 0.9480 \\
% & \textsc{Finetune-Based} (MS-MARCO) & ranking & 0.7984\\
\midrule
%Ours & \textsc{Prompt-Based-LabelCond} (ESCI) & classification & 0.9445\\
Ours & \textsc{Finetune-Based-LabelCond} (ESCI) & classification & 0.9151	& 0.9513 & 0.9544\\
%  & \textsc{Finetune-Based-AllLabes} (MS-MARCO) & classification & 0.8318\\
\bottomrule
\end{tabular}
}
\caption{Results on HomeDepot dataset.
}
\label{tab:homedepot_results}
\vspace{-6mm}
\end{table*}

%% file: tables/qgen_example.tex
\begin{table*}
\resizebox{\textwidth}{!}{
\begin{tabular}{llll}
\toprule
\textbf{Desired Labels} & \textbf{\textsc{Prompt-Based-LabelCond}} & \textbf{\textsc{Finetune-Based-LabelCond}} & \textbf{\textsc{Finetune-Based-LabelCond} } \\
& \textbf{(ESCI)} & \textbf{(ESCI)} & \textbf{(MS-MARCO)} \\
\midrule
 Label: E & bed frame with storage plans free &  wood bed frame & -   \\
 Label: S & acacia wood bed & king bed frame with drawers & - \\
 Label: C & how much does a queen size mattress cost & adjustable bed frame king & - \\
 Label: I & what is the best platform bed frame & headboard with lights and frame & - \\
\midrule
 Label: Relevant & - & - &  what kind of wood is used for bed frames  \\
 Label: Irrelevant & - & - & what is a king size bed frame \\
 \midrule
 \midrule
 Gold Label: Exact & \multicolumn{3}{l}{hardwood beds} \\
 Gold Label: Partial & \multicolumn{3}{l}{geralyn upholstered storage platform bed, floating bed, beds that have leds} \\
 Gold Label: Irrelevant & \multicolumn{3}{l}{jordanna solid wood rocking} \\
\bottomrule
\end{tabular}
}
\caption{Some examples of generated queries from different QGen models for 
product ``solid wood platform bed good , deep sleep can be quite difficult to have in this busy age . fortunately , there ’ s an antidote to such a problem : a nice , quality bed frame like the acacia kaylin . solidly constructed from acacia wood , this bed frame will stand the test of time and is fit to rest your shoulders on for years and years . its sleek , natural wood grain appearance provides a pleasant aesthetic to adorn any bedroom , acting both as a decorative piece as well as a place to give comfort after a hard day of work . our bed frame is designed to give ample under-bed space for easy cleaning and other usages , with a headboard attached to further express the craftiness . it can be used with other accessories such as a nightstand or bookcase headboard and is compatible with many types of mattresses including memory foam , spring , or hybrid ones . there ’ s nowhere better to relax than your own home , and with this bed frame that feeling of homeliness will even be more emphasized . rest comfortably and in style .
''. For reference, we also provide the gold queries from the WANDS test data for the reader.}
\label{tab:wands_examples_1}
\vspace{-6mm}
\end{table*}

%% file: 04-results.tex
\section{Results and Discussion} \label{sec:results}
In this section, we present the results of two major QGen families (finetune-based vs prompt-based) comparing them with cross-domain transfer learning approach. 
Since WANDS is a more recent and challenging dataset in comparison to HomeDepot\footnote{We refer the reader to \citet{wands} for more information. }, we focus on WANDS for our discussion. 
We report  results for WANDS in \autoref{tab:wands_results} and for HomeDepot in \autoref{tab:homedepot_results}.
Here are our main findings:

\textbf{Zero-shot Transfer Learning wins over any QGen!}
Overall, we find that  zero-shot transfer learning outperform all QGen approaches, both vanilla and label-conditioned.
This is unlike what existing works such as \textsc{InPars} and \textsc{Promptagator} where QGen approaches give the best downstream performance.
This could be attributed to the difficulty of the downstream task, which in this case is a nuanced relevance prediction task, while the existing works focus on binary relevance which is much simpler.

\textbf{Label-conditioned QGen wins over vanilla QGen!}
Within the QGen approaches, we find that our adaptation of conditioning on all relevance labels outperforms the vanilla versions which do not.
From the results of \textsc{Finetune-Based} and \textsc{Finetune-Based-LabelCond} trained on MS-MARCO, we find that exposing the QGen models to all labels (in the case of MS-MARCO they are binary) performs better by +3.3 NDCG@10 points.
Therefore, we finetune with all labels on a related dataset (ESCI) for WANDS and find that it outperforms even the MS-MARCO-based QGen models.
For prompt-based QGen models, we find that its label-conditioned counterpart underperforms its vaniila variant.
However, the prompt-based vanilla variant is far behind (-8.3 NDCG@10 points) the finetune-based vanilla variant to begin with.

\textbf{In-domain training is important!}
We find that for both transfer learning and QGen approaches, transferring from a related domain is important in downstream performance.
For instance, within the transfer learning models, the model trained on ESCI (zero-shot (ESCI)) gives the best downstream performance, even 
outperforming the model trained on MS-MARCO (zero-shot (MS-MARCO)), which is trained on nearly 10 times larger training data than ESCI.
This again emphasizes that having a related dataset to transfer from is essential for downstream performance, similar to \citet{gururangan-etal-2020-dont}.
Similarly, within QGen approaches, the label-conditioned model trained on ESCI (\text{Finetune-Based-LabelCond (ESCI)}) outperforms its MS-MARCO counterpart.
Clearly, relatedness of the target dataset to the training dataset is also important for the QGen model training.

\input{tables/qgen_data}
Below we discuss the probable reasons for the shortcomings of QGen approaches.
We inspect three QGen models which have been trained with all labels, namely, \textsc{Prompt-Based-LabelCond (ESCI)}, \textsc{Finetune-Based-LabelCond (ESCI)} and \textsc{Finetune-Based-LabelCond (MS-MARCO)}, for the number of duplicate queries generated by the model.
Specifically, a duplicate query here refers to the QGen model producing the same query across different relevance labels for the same product.
In \autoref{tab:wands_results_duplicate} we report the results for WANDS.
As you recall, for each of the WANDS 42,994 products, the QGen models trained on ESCI, generated 171,976 queries, one for each of the four relevance labels.
For QGen models using MSMARCO, we generate 85,988 queries, one of each of the two relevance classes.
In \autoref{tab:wands_results_duplicate} we find that the \textsc{Finetune-Based-LabelCond (ESCI)} QGen model produces duplicate queries for 81\% of the products, which suggests that simply prepending label information in the input context is insufficient for the model to learn how to generate discriminative queries.
We would also like to highlight the fact that this is happening despite exposing the QGen model to the entire ESCI training data which is 1.6 million examples, of which only 5 of the 1.1 million products had duplicate queries.
In \autoref{tab:qgen_data} we report the distribution of generated queries across different labels, after applying the filtration step (described in \autoref{sec:setup}) where we remove the duplicate queries.
Clearly, noise in the synthetic queries causes errors in the subsequent downstream models.
Interestingly, despite  \textsc{Prompt-Based-LabelCond (ESCI)} and \textsc{Finetune-Based-LabelCond (MS MARCO)} models having more number of valid queries, they still underperform \textsc{Finetune-Based-LabelCond (ESCI)}.
\input{tables/wands_duplicate_queries}

The reason why \textsc{Finetune-Based-LabelCond (MS-MARCO)} is underperforming its ESCI counterpart could be attributed to a) the difference in domain and, b) the style of queries.
For instance, queries from MS-MARCO-trained-QGen models are more formal \emph{what-}style questions, while queries from ESCI-trained-QGen models are more informal and similar in style to the gold queries.
Although,  \textsc{Prompt-Based-LabelCond (ESCI)} has far fewer duplicate queries it severely underperforms probably because of poor overall quality.
In \autoref{tab:wands_examples_1} we present the generated queries from 
different QGen model for a product.
We also provide the user-issued or gold query from the WANDS test set for the same product.\footnote{Note that all products in the test set do not have queries for each relevance label, we simple sampled from those products which have for qualitative evaluation purposes.}
For \textsc{Prompt-Based-LabelCond (ESCI)} we  see that the query for the highest relevance label i.e. `E' focuses on the entity bed frame with free storage plans, while from the product description we know that it is mainly about a bed frame which is made from acacia wood and additionally has storage.
Nowhere does the product talk about storage plans.
In fact, the query for the next relevance label `S' is more relevant than the one for `E'.
Clearly, exposing the models to only 8 examples, as proposed by \textsc{promptagator} \cite{dai2022promptagator}  is insufficient, in comparison to the 1.6 million examples used by \textsc{Finetune-Based-LabelCond (ESCI)}, especially for the WANDS dataset.
On the other hand, \textsc{promptagator} work had found that exposing the models to only 8 task-specific examples for QGen had outperformed finetuned models which were trained on $\mathcal{O}(100k)$ MS-MARCO examples.
We would like to note that the \citet{dai2022promptagator} also apply an additional consistency filtration step to the generated queries, wherein they only retain those queries which are answerable from the passage from which it was generated.
They find that that adding this round-trip consistency adds 2.5 points (avg.) but for smaller datasets it negatively impacts the downstream performance.
Therefore, we experimented with round-trip consistency for the \textsc{Finetune-Based-LabelCond (ESCI)} model for WANDS, which is the best among all QGen variants.
Specifically, we use the downstream relevance prediction model trained on ESCI (i.e. the model used for \textbf{zero-shot} transfer learning) and re-label the generated queries.\footnote{\citet{dai2022promptagator} use the downstream model trained on synthetic data instead we use a model trained on good quality ESCI data}
We first find that the predicted label of 49\% of the generated queries do not match the label which was used to generate the query (i.e. the desired label).
We then use the predicted label as the final label for that query and train a downstream model as before.
We find this results in only +1 point improvement.\footnote{Given that, for WANDS  in \autoref{tab:wands_results} the \textsc{Prompt-Based-LabelCond (ESCI)} is almost 21 points behind its finetuned counterparts, we did not apply this additional step which would require additional model training.}

This highlights that even though QGen techniques offer a promising solution for adapting models to new domains, they need further investigation and analyses to make them more effective across different tasks.

%% file: tables/qgen_data.tex
\begin{table*}
\resizebox{0.75\textwidth}{!}{
\begin{tabular}{lll}
\toprule
\textbf{QGen Model} & \textbf{Desired Label} & \textbf{Generated Query Distribution} \\
\midrule
\textsc{Prompt-Based-LabelCond} (ESCI) & Label: E & 32816 \\
& Label: S & 33271\\
& Label: C & 34998\\
& Label: I & 33455\\
& All & 134540\\
\midrule
\textsc{Finetune-Based-LabelCond} (ESCI) & Label: E & 37573\\
& Label: S & 28969\\
& Label: C & 26986\\
& Label: I &  23651\\
& All & 117179\\
\midrule
 \textsc{Finetune-Based-LabelCond} (MS-MARCO) & Label: Relevant & 42790 \\
 & Label: Irrelevant & 40933 \\
 & All & 83723 \\
\bottomrule
\end{tabular}
}
\caption{Total number of generated queries (post filtering) for WANDS.
Total number or products: 42994.
}
\label{tab:qgen_data}
\end{table*}

%defaultdict(<function <lambda> at 0x7ff76686f880>, {'Relevant': 42790, 'Irrelevant': 40933})
%83723

%defaultdict(<function <lambda> at 0x7f955be71240>, {'C': 26986, 'E': 37573, 'I': 23651, 'S': 28969})
%117179

%defaultdict(<function <lambda> at 0x7f3779727520>, {'S': 33271, 'E': 32816, 'C': 34998, 'I': 33455})
%134540

%% file: tables/wands_duplicate_queries.tex
\begin{table*}
\resizebox{\textwidth}{!}{
\begin{tabular}{llll}
\toprule
\textbf{Number of Products} & \textbf{\textsc{Prompt-Based-AllLabels}}  & \textbf{\textsc{Finetune-Based-AllLabels}} & \textbf{\textsc{Finetune-Based-AllLabels}}  \\
& \textbf{(ESCI)} & \textbf{(ESCI)} & \textbf{(MS-MARCO)} \\
\midrule
at least 1 duplicate  & 2825 &  35126 & 2335 \\
duplicate query for E and S & 543 & 13805 & NA \\
duplicate query for S and C & 595 & 14105 & NA \\
duplicate query for C and I & 571 & 15687 & NA \\
\bottomrule
\end{tabular}
}
\caption{Measuring for how many times does the QGen model produce the same or duplicate query across different relevance labels for the same product.
For MS-MARCO-QGen model, remember we only had binary relevance labels, hence rows 2--4 are not applicable for this model.
Total number or products: 42994.
}
\label{tab:wands_results_duplicate}
\vspace{-6mm}
\end{table*}

%% file: 05-relatedwork.tex
\section{Related Work}
Synthetic Question Generation has come a long way from relying on simple but rigid heuristics \cite{Smith2011AutomaticFQ}  to using neural-network approaches, specifically seq-to-seq model \cite{serban-etal-2016-generating, Zhou2017NeuralQG, du-etal-2017-learning, du-cardie-2018-harvesting}, to now even leveraging large language models (LLMs) through prompting \cite{dai2022promptagator}. 
Much of the work in this area has focussed on question generation in the context of QA systems.
Below we describe some of the representative works in this area.

\paragraph{QGen for QA}
Pre-transformer era had seq-to-seq models trained with attention  to read an input sentence and generate a question with respect to an answer which is contained in that sentence e.g. for factiod QA \cite{Zhou2017NeuralQG, du-etal-2017-learning} 
% In addition to the sentence input and the gold answer, some combine handcrafted lexical features with answer position information to generate the question, for which they train on the popular SQuaD \cite{rajpurkar2016squad} dataset.
% However, they investigate the generated question-sentence-answer triples using BLEU metric but do not explore its effect on improving downstream QA systems.
% \citet{du-etal-2017-learning} although do not use any handcrafted features but they also limit their evaluation of generated factoid questions using only BLEU and not on any downstream tasks, similar to \citet{Zhou2017NeuralQG}.
\citet{du-cardie-2018-harvesting} go beyond using single sentence context (as \citet{du-etal-2017-learning} note that 30\% of SQuaD questions span answers beyond single sentence) for generating questions.
Transformers \cite{vaswani2017attention} changed the game subsequently with their power of attention to refer to specific parts of text -- the QGen models have further improved.
For instance, \citet{lopez2020transformer} use a GPT-2 \cite{radford2019language} language model to train a question-generation model using the passage as input.
They also train an answer-aware variant where they mark start and end of the answer span with special tokens in the context.
However, they find the answer-aware variant to be under-performing for question generation (in terms of BLEU metric) than the answer-unaware model.
They hypothesize that this is because  there is no explicit mechanism to inform the model on how to use the answer information, somewhat similar to what we find in our label-conditioned models as well where they seem to not use the label information effectively.
\citet{unlu-menevse-etal-2022-framework} explore question-generation for Spoken QA task.
More recently, \citet{ko-etal-2020-inquisitive, chakrabarty-etal-2022-consistent, cao-wang-2021-controllable} propose approaches to generate more open-ended questions, whose answers often span multiple sentences and could be long-form.
\citet{cao-wang-2021-controllable} create a question-type ontology to guide the model to generate a particular type of question.
They essentially concatenate the question-type with the multi-sentence input to generate the question.
In the hope of controlling the question generation, they train it jointly with question focus prediction which uses semantic graphs.
In principle the question focus and label conditioning are related as in our case, question focus is the conditioned label, however, 
their main goal of work is to generate questions which are  diverse and illicit complex reasoning or curiosity \cite{ko-etal-2020-inquisitive}.
It is not evaluated on improving any downstream tasks.

\paragraph{Label-Conditioned QGen}
Some previous work have looked at label conditioning in QGen models for classificication tasks.
\citet{kumar-etal-2020-data, yang2020generative} find that pre-pending class-labels to input text is quite effective in class-conditional test generation and thereby data augmentation.
They show the effectiveness of this approach for classification tasks (e.g. SST-2 with binary relevance, SNIPS with 7 intents, and TREC with six-classes, SNLI, commonsense reasoning) across different pretrained LMs including auto-encoder LM (BERT \citet{devlin2018bert}), auto-regressive LM (GPT-2 \citet{radford2019language}) and pretrained seq-2-seq LM (BART \cite{lewis-etal-2020-bart}).
In this work, we look at fine-grained relevance prediction, where the task is difficult in that the multiple classes have an inherent ordering, and therefore  it is harder for QGen models to produce discrimnative queries across such fine-grained labels.

% \paragraph{Issue of}
% \citet{yang2020generative} use additional filtering techniques (e.g. influence functions) to filter noisy and less-informative synthetic data.
% As expected, with the state-of-the-art LLMs, the fluency of the generated text has improved, however \citet{kumar-etal-2020-data, he2022generate} find that often the generated text conditioned on a given label is not faithfull to it. 
% \citet{he2022generate} in fact find that one can discard the conditioned labels and instead use soft-labels from a teacher model. 

%% file: 05-conclusion.tex
\section{Limitations and Next Steps}
From the above results, it is apparent that QGen approaches, although offering a promising direction especially for zero-shot settings, need considerable work to outperform transfer learning.
Clearly, simply adding label information in the input context does not provide a sufficient signal for the model to generate discriminative queries.
We need to explicitly enforce this signal  throughout the QGen training process.
In this work, we only generate one query, but using beam search we could generate multiple queries for a given product-label combination, resulting in a diverse collection.
Another challenge in working with QGen approaches is that the typical strategy for evaluating the synthetic data is to evaluate it on a downstream task, requiring two additional steps after training a QGen model: applying the QGen model for generating queries and then  training a downstream task model, to understand the effect of synthetic data.
So if a researcher wanted to experiment with multiple QGen models, they would have to run three times the number of experiments to understand which QGen model is the best one, which is a waste of resources and time.
This means that we need to come up with an intrinsic evaluation metric that correlates well the downstream task performance.
Our next steps are focused on addressing these issues.

%% file: appendix.tex
%\appendix
%\section{Appendix}

\input{tables/example_data}

\input{tables/label_stats}

%% file: tables/example_data.tex
\begin{table*}
\resizebox{0.8\textwidth}{!}{
\begin{tabular}{p{.7in}p{1in}p{.4in}p{2.3in}}
\toprule
\textbf{Dataset} & \textbf{Query} & \textbf{Label} & \textbf{Product Data} \\ 
\midrule

ESCI &
calculator texas instruments &
E &
Title: Texas Instruments TI-84 Plus CE Color Graphing Calculator, Black 7.5 Inch
\newline
Product Bullet Point: High-resolution, full-color backlit display Rechargeable battery ...  \\
\midrule

MS-MARCO &
what is the role of mast cells in inflammation &
0 &
Document: Cells of the Immune System Cell types with critical 
roles in adaptive immunity are antigen-presenting cells ... \\
\midrule

WANDS &
salon chair &
Exact &
Title: 21.7 '' w waiting room chair with ...
\newline
Description: this is a salon chair , barber chair for a hairstylist ... \\
\midrule

Home Depot &
jeldwen 24 inch bifold doors &
2.33 &
Title: JELD-WEN Smooth 2-Panel Arch Top Hollow Core Molded Interior Closet ...
\newline
Description: The 2-Panel Arch Top Interior door from JELD-WEN has a classic design ... \\

\bottomrule
\end{tabular}
}
\caption{Examples from the ESCI, MS-MARCO, WANDS and HomeDepot datasets.}

\label{tab:qgendatasets}

\end{table*}

%% file: tables/label_stats.tex
% Table for Label Stats
% Please refer to cl/527731285.

\begin{table*}[t!]
\resizebox{0.5\textwidth}{!}{
\begin{tabular}{lrrrr}
\toprule
\textbf{Dataset} & \textbf{Label} & \textbf{Train} & \textbf{Dev} & \textbf{Test} \\ 
\midrule
ESCI & E & 1,093,105 & 13,948 & - \\
     & S & 351,975 & 4,357 & - \\
     & C & 45,662 & 577 & - \\
     & I & 160,870 & 2,024 & - \\
\midrule
MS-MARCO & 0 & 18,646,285 & 13,960 & - \\
         & 1 & 18,646,285 & 6,492 & - \\
\midrule
WANDS & Exact & - & - & 25,614 \\
      & Irrelevant & - & - & 61,201 \\
      & Partial & - & - & 146,633 \\
\midrule
Home Depot & 1st Quartile & - & - & 5,095 \\
           & 2nd Quartile & - & - & 6,773 \\
           & 3rd Quartile & - & - & 27,704 \\
           & 4th Quartile & - & - & 34,217 \\
\bottomrule
\end{tabular}
}
\caption{Labels Distribution for the ESCI, MS-MARCO, WANDS and HomeDepot datasets. We used ESCI and MS-MARCO only for training (so no test set stats shown) and we used WANDS and Home Depot only for testing as a downstream task (so only test set stats are shown).
}

\label{tab:qgenlabels}

\end{table*}